\documentclass[sigconf]{acmart}

\AtBeginDocument{%
  \providecommand\BibTeX{{%
    \normalfont B\kern-0.5em{\scshape i\kern-0.25em b}\kern-0.8em\TeX}}}

\setcopyright{acmcopyright}

\copyrightyear{2020} 
\acmYear{2020} 
\setcopyright{acmlicensed}\acmConference[KDD '20]{Proceedings of the 26th ACM SIGKDD Conference on Knowledge Discovery and Data Mining}{August 23--27, 2020}{Virtual Event, CA, USA}
\acmBooktitle{Proceedings of the 26th ACM SIGKDD Conference on Knowledge Discovery and Data Mining (KDD '20), August 23--27, 2020, Virtual Event, CA, USA}
\acmPrice{15.00}
\acmDOI{10.1145/3394486.3403387}
\acmISBN{978-1-4503-7998-4/20/08}

\begin{document}

\title{Intelligent Exploration for User Interface Modules of Mobile App with Collective Learning}

\author{Jingbo Zhou$^{1,3*}$,
        Zhenwei Tang$^{1,6}$,
        Min Zhao$^2$,
        Xiang Ge$^2$,
        Fuzhen Zhuang$^{4,5}$
        }
\author{
        Meng Zhou$^{1,7}$,
        Liming Zou$^2$,
        Chenglei Yang$^8$,
        Hui Xiong$^{9*}$
        }
\thanks{$^*$Jingbo Zhou and Hui Xiong are corresponding authors (zhoujingbo@baidu.com, hxiong@rutgers.edu).}
\affiliation{$^1$Business Intelligence Lab, Baidu Research, $^2$Baidu TPG User Experience Department, China}
\affiliation{$^{3}$National Engineering Laboratory of Deep Learning Technology and Application, China}
\affiliation{$^{4}$Key Lab of IIP of Chinese Academy of Sciences (CAS), Institute of Computing Technology, CAS, Beijing}
\affiliation{$^{5}$University of Chinese Academy of Sciences, Beijing, $^{6}$Beijing University of Posts and Telecommunications}
\affiliation{$^{7}$Peking University,  $^{8}$Shandong University, $^{9}$Rutgers University}
\renewcommand{\shortauthors}{Zhou, et al.}
\fancyhead{}
\begin{abstract}
A mobile app interface usually consists of a set of user interface modules. How to properly design these user interface modules is vital to achieving  user satisfaction for a mobile app.
However, there are few methods to determine design variables for user interface modules except for relying on the judgment of designers. Usually, a laborious post-processing step is necessary to verify the key change of each design variable. Therefore, there is a only very limited amount of design solutions that can be tested. It is time-consuming and almost impossible to figure out the best design solutions as there are many modules. 
To this end, we introduce FEELER, a framework to fast and intelligently explore design solutions of user interface modules with a collective machine learning approach.
FEELER can help designers quantitatively measure the preference score of different design solutions, 
aiming to facilitate the designers to conveniently and quickly adjust user interface module.
We conducted extensive experimental evaluations on two real-life datasets to demonstrate its applicability in real-life cases of user interface module design in the Baidu App, which is one of the most popular mobile apps in China.  
\end{abstract}

\keywords{User interface exploration, user interface design, collective learning}


\maketitle

\section{Introduction}
\begin{figure} 
    \includegraphics[width=0.8\columnwidth]{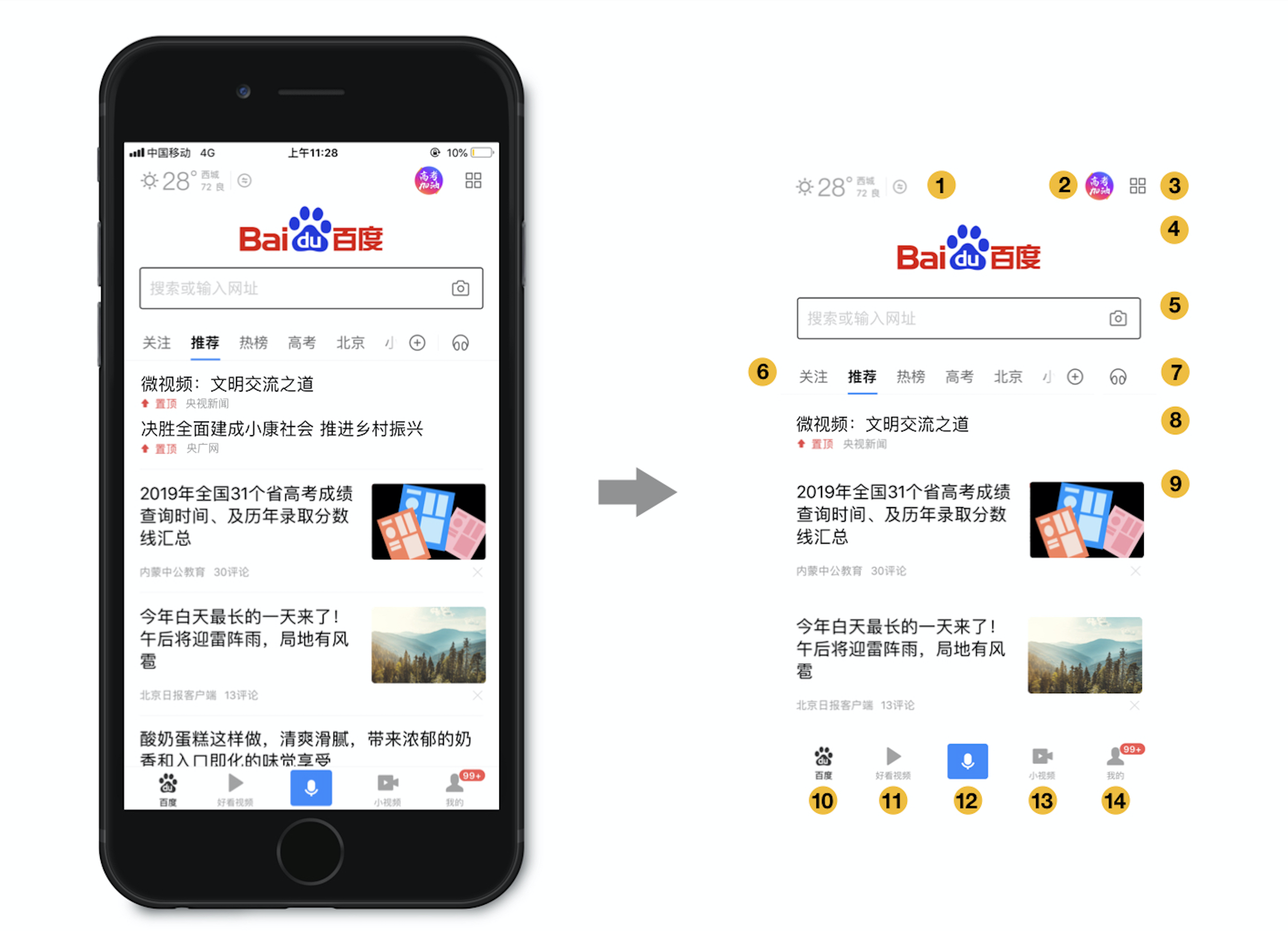}
    \vspace{-3mm}
    \caption{The user interface modules of the Baidu App. There are 14 user interface modules, where Search Box (No. 5) and News Feed (No. 8 and No. 9) are two examples.}~\label{fig:baidu_uims}
\vspace{-6mm}
\end{figure}

A user interface of a mobile app can disassemble into different user interface modules. 
In Figure \ref{fig:baidu_uims}, we illustrate the important modules on the user interface of the Baidu App which has more than 200 million daily active users and is one of the most popular mobile apps in China. As we can see from the right side of Figure \ref{fig:baidu_uims}, there are mainly 14 modules, and most of them play an important role in the functionality of the app, such as the Search Box (No. 5) module and the News Feed (No.8 and No. 9) module. 
Finding the best design solutions to such modules is critical to improve user satisfaction of the mobile app.


The design of a mobile app's user interface is usually conducted in two levels.  At a lower level, designers will provide the design solution of each user interface module. As shown in Figure \ref{fig:searchbox_var}, a user interface module (i.e. Search Box) usually has several key design variables. Note that all the design variables here refer to visual appearance variables of a user interface module.
With varying such design variables, we can get different design solutions for the user interface module.  Figure \ref{fig:searchbox_uim_instance} illustrates different design solutions of the Search Box module.
At a higher level, the designers combine these modules into a whole interface. While the whole interface is some kind of fixed, the modules at the lower level are always adjusted by designers due to several reasons, such as changing styles in different holidays, adding temporary modules and revising important modules. 
Hence, designers are usually exhausted to adjust the design solutions of user interface modules. 

In this paper we investigate how to intelligently explore the best design solutions for user interface modules, aiming to build a predictive model that can assess the preference score of a given user interface module. The predictive module should also be able to quantitatively measure the preference score of different design solutions, 
and analyze the correlations among variables. In this way, the model can help designers conveniently and quickly adjust user interface modules. Though how to design the whole interface at a higher level is also a challenging research topic, it is beyond the scope of this paper.

However, to the best of our knowledge, there are few existing studies to help identify a better design solution of user interface modules. 
 There is a combinatorial explosion for enumerating all possible design solutions. Thus, most of the design variables of an interface module are determined by designers according to their judgement and personal preference. Traditionally, designers would come up with a couple of different designs and then verify whether users like them or not by a post-processing step, using online A/B test or offline evaluation. Such a post-processing step is usually time-consuming and requires high labor costs. Hence, only a few design solutions of the user interface module can be tested and properly evaluated. In this way, it is almost impossible to find out the best design solutions. In recent years, there are some works about evaluating the user experience of a product \cite{seguin2019triptech}, and the friendliness of the machine learning interface \cite{kayacik2019identifying}. But all of them are survey-based method without using machine learning technology. Machine learning methods have been used to tappability \cite{swearngin2019modeling} or accessibility \cite{guo2019statelens} problems, which usually makes prediction based on existing screen without attempting to adjust the design solution. To the best of our knowledge,  there is no existing study to employ machine learning to explore better design solutions of user interface modules.

In this paper, we propose FEELER, a method of Intelligent Exploration for User Inter\underline{f}ac\underline{e} Modul\underline{e} of Mobile App with  Col\underline{l}ective L\underline{e}a\underline{r}ning. 
The core of FEELER is to use a two-stage collective learning method to build a predictive model based on the user feedback for interface modules collected by multiple rounds in a crowdsourcing platform. 


The first stage of FEELER is to build a proactive model with active learning. The proactive model has an iterative optimization process to find the best values of a predictive function with minimized cost, using a crowdsourcing platform to invite participants to rate their preference of different design solutions. A challenge of this stage is how to manage the exploitation versus exploration trade-off, i.e. the ``exploitation'' of the design solutions that has the highest expected preference scores and ``exploration'' to get more diverse design solutions. Hence, an acquisition function is defined in the proactive model to guide the exploration of design solutions in each round with balancing the exploitation versus exploration trade-off. 

The second stage of FEELER is a comparison-tuning model which can further improve the predictive performance for design solutions upon the predictive ability of the proactive model. The comparison-tuning model is motivated by the following two insights. First of all, our major concern is how to distinguish the best design solutions from the good ones (the bad design solutions are obvious and not useful). Second, the participants usually can only differentiate which solutions are better after comparing them. Therefore, in this stage, we generate pairs of design solutions based on the ones returned by the proactive model and then invite participants to rate which solution is better for each pair. The comparison-tuning model is optimized based such labeled pairwise comparison data. 

In addition, FEELER also provides a mechanism to quantitatively analyze the design variables for each user interface module, and the correlation among design variables. 
In this way, designers can adjust the design variables for each user interface module while being aware of the module preference scores.  In this perspective, another benefit of FEELER is to bring the quantitative analysis methodology for user interface design. 

At last, we conduct extensive experiments on two important user interface modules, Search Box and News Feed, of the Baidu App to show the effectiveness of FEELER over baselines. We also conduct an in-depth analysis of the design variables upon the predictive models of FEELER, to demonstrate how such results can help designers. FEELER has been used to guide the design of the Baidu App in practice.  
The findings, limitations and further research opportunities are also discussed.

We summarize the contributions of this paper as follows:
\begin{itemize}
    \item We are the first to study the exploration of user interface modules with a machine learning method. Our research sheds some light on a new user interface design paradigm with machine learning methodology. 
    
    \item We propose a method, called FEELER, for intelligent exploration for user interface modules based on a multiple round crowdsourcing process. FEELER has two major stages which are to build a proactive model and a comparison-tuning model respectively.
    
    \item We conduct extensive experimental evaluations and in-depth model analysis on two real-life datasets from Search Box and New Feed of the Baidu App, to demonstrate the effectiveness and utility of FEELER. 
    
\end{itemize}

\section{Overview}
\subsection{Preliminaries}


\begin{figure}
   \includegraphics[width=0.8\columnwidth]{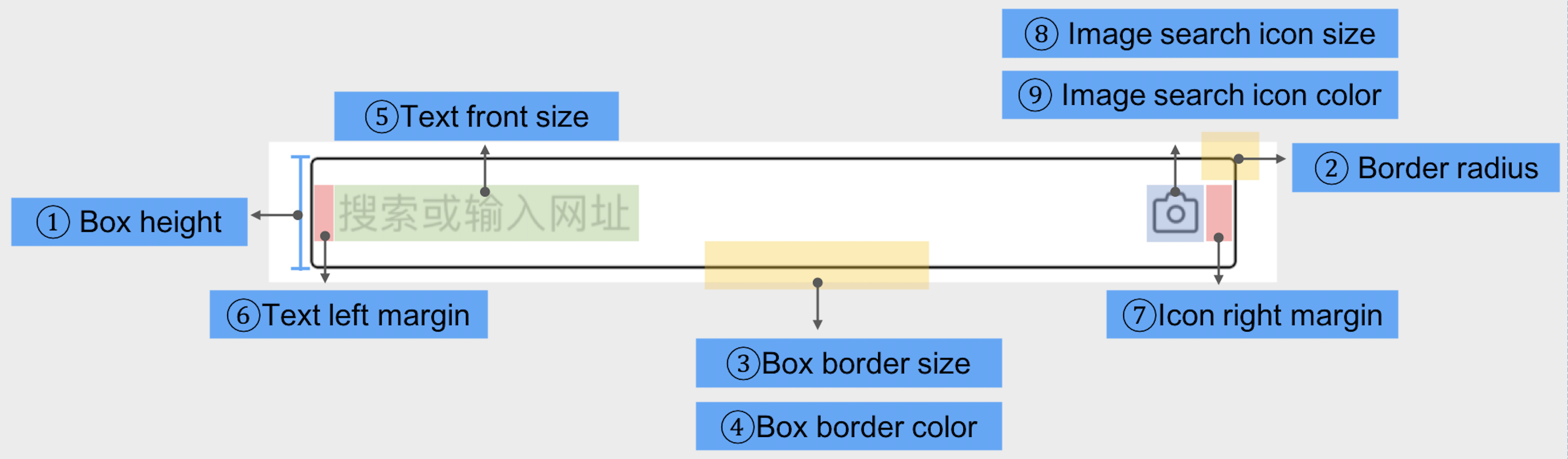}
\vspace{-2mm}
   \caption{Design variables of Search Box}~\label{fig:searchbox_var}
\vspace{-6mm}
\end{figure}

\begin{figure}
    \includegraphics[width=1\columnwidth]{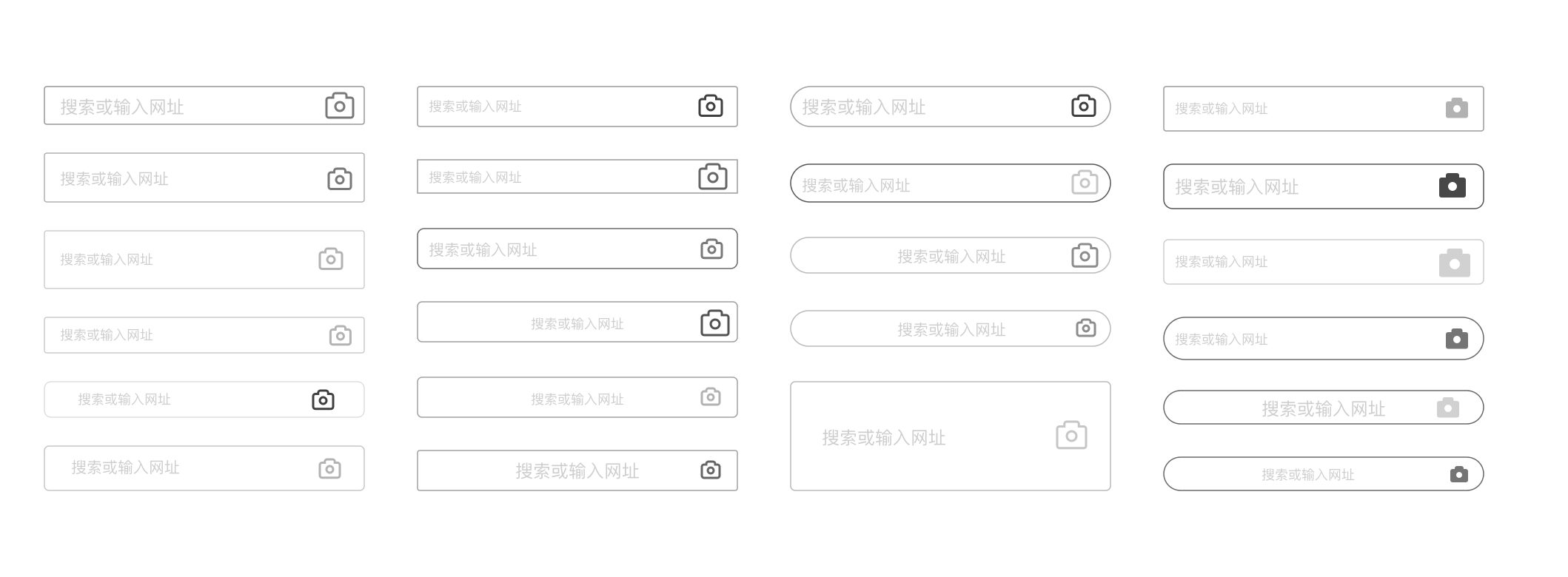}
    \vspace{-6mm}
    \caption{Different design solutions of Search Box}~\label{fig:searchbox_uim_instance}
\vspace{-8mm}
\end{figure}

Here we introduce preliminaries using over this paper. As described in the introduction section, each user interface of a mobile App has several modules. A module of a user interface usually has a set of design variables. We name such a set of parameters for a module as a design variable vector $\vec{x}$. 
All the design variable vectors of a module belong to a predefined domain $\mathcal{X}$, i.e. $\vec{x} \in \mathcal{X}$. Each design variable vector defines a design solution of a module.  As we can see from Figure \ref{fig:searchbox_var}, the design variables of Search Box on the Baidu App include the color, thickness, height of the box, the font size of the default search text, and so on. 
Figure \ref{fig:searchbox_uim_instance} illustrates different design solutions of the Search Box in the Baidu App with varying design variable vectors.

It is not necessary to add all the design variables into the model since some design variables can be easily determined or are not important factors. Besides, a large magnitude of variables adds to the difficulty of effectively and efficiently constructing the model. 
In FEELER, the key design variables are selected after discussion with designers and user experience researchers together. The design variables that considered to be very important to user experience based on judgments of designers, or variables that designers were eager to explore more, were given high priority to be included. In our system, there are 9 design variables for Search Box, and 8 design variables for News Feed. 

Given an oracle model $\hat{\psi}(\cdot)$ returning the general user preference of a design solution, the exploration of user interface module can be considered as a process to find the best design variable vector $\vec{x}^*$ according to the model $\hat{\psi}(\cdot)$, i.e. $\arg{max}_{x\in \mathcal{X}} \hat{\psi}(x)$. 
Usually, the designers adjust the design variables iteratively to find the best design solutions for each user interface module. In this paper, we aim to build a surrogate model $\psi(\cdot)$ to approximate the oracle function $\hat{\psi}(\cdot)$.  Note that the objective of FEELER is to explore the best design solution, instead of approximating $\hat{\psi}(x)$ perfectly. Therefore, we require the surrogate model $\psi(\cdot)$ to be accurate when the design variable vector nearby the ones of the best solutions.  
In other words, it is not so useful to make accurate user preference prediction when the design solution is far from the best ones.

In FEELER we choose Gaussian Processes (GPs) as the base model to approximate the oracle model $\hat{\psi}(\cdot)$.  GP is a rich and flexible class of non-parametric statistical models over function spaces \cite{williams2006gaussian} with a wide range of applications \cite{zhou2015smiler}. The reasons to select GPs can be explained from three perspectives. At first, the output of GPs is probabilistic so that we can obtain confidence intervals for each prediction by GPs. Such confidence intervals are very important information for designers. Second,  GPs provide a natural mechanism to define the acquisition function for active learning to balance the exploitation versus exploration trade-off. Third, GPs can be optimized by the pairwise comparison data. Instead of directly giving a preference score, the user can more precisely express their preference by comparing a pair of design solutions. GPs can utilize such pairwise comparison data for model optimization. We will further explain the second and the third advantages of GPs in Section \ref{sec:activelearing} and Section\ref{sec:pl} respectively.

 Here we give a brief introduction about GPs \cite{williams2006gaussian}. Given a set of labeled training data $D={(x_i,y_i)}$ where $x_i$ is a design variable vector and $y_i$ is labeled preference score, a model $\psi(\cdot)$ aims to predict the score of a test design variable vector $\vec{x}_{t}$.  GPs assume the $\psi$ is drawn from a GP prior that $\mathcal{P}(\psi) \sim \mathcal{N}(0, \mathbf{K})$ where $\mathcal{N}(0, \mathbf{K})$ is Gaussian distribution with mean is zero and covariance matrix is $\mathbf{K}$. Note that zero-mean assumption is for simplicity which is not a drastic limitation, since the mean of the posterior process is not confined to be zero \cite{williams2006gaussian}. The covariance matrix $\mathbf{K}$ is also called the Gram matrix \cite{preoctiuc2013temporal} whose elements $k_{i,j}$ are defined by a kernel function over a pair of training instances, i.e. $k_{ij}=\kappa(\vec{x}_i,\vec{x}_j)$. In our case study, we use the popular Radial Basis Function (RBF) kernel, where $\kappa(\vec{x}_i,\vec{x}_j)=exp(-\frac{\|\vec{x}_i-\vec{x}_j\|}{2\Delta^2})$. The posterior probability of the test vector $\vec{x}_{t}$ after observing the training data $D$ is:
 \begin{align}
\mathcal{P}(\psi(x_t)|D) = \int \mathcal{P}(\psi(x_t)|\psi)\mathcal{P}(\psi|D)\mathrm{d}\psi.
 \end{align}
 The posterior predictive probability distribution  of $\mathcal{P}(\psi(x_t)|D)$ can be solved analytically
 which is also a Gaussian distribution  $ \mathcal{P}(\psi(x_t)|D) = \mathcal{N}(u(\vec{x}_{t}), \delta^2(\vec{x}_{t}))$ with:
 \begin{align}
 u(\vec{x}_{t}) &= \mathbf{k}^T\mathbf{K}^{-1}Y \label{eqn:gp_mean},\\
 \delta^2(\vec{x}_{t}) &= \kappa(\vec{x}_{t}, \vec{x}_{t}) - \mathbf{k}^T\mathbf{K}^{-1}\mathbf{k}, \label{eqn:gp_var}
 \end{align}
 where $\mathbf{k}$ is kernel vector evaluated between the test vector $\vec{x}_t$ and all training instances, i.e. $\mathbf{k} = [\kappa(\vec{x}_t,\vec{x}_1), \kappa(\vec{x}_t,\vec{x}_2), ....]^T$, and $Y=[y_1,y_2,...]^T$.

\subsection{Framework of FEELER}
\begin{figure*}
   \includegraphics[width=1.8\columnwidth]{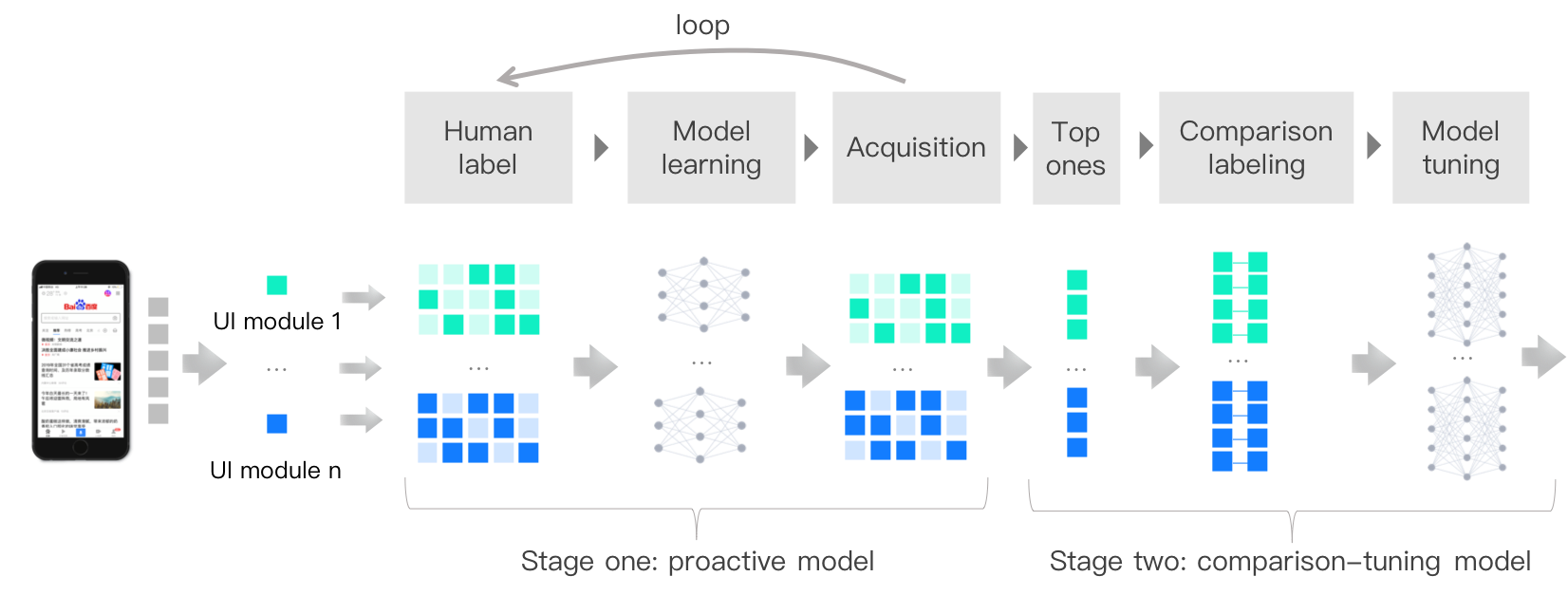}
   \vspace{-4mm}
    \caption{Overview of FEELER. The first stage of FEELER constructs the proactive model, and the second stage of FEELER builds the comparison-tuning model.}~\label{fig:overview}
\end{figure*}

We propose a two-stage framework to approximate the oracle model $\hat{\psi}(\cdot)$ leveraging the collectively labeled data by crowdsourcing. An illustration of the FEELER is shown in Figure \ref{fig:overview}. The first stage of FEELER is called proactive model with an iterative optimization process, and the second stage is named as comparison-tuning model which is a fine-tuning prediction model by pairwise comparison. 

 In the first stage, we train a proactive model with crowdsourcing labeled data.  At first, we generate a set of design solutions for a given module and then recruit many participants to rate the design solutions. Here we use Bayesian-based active learning \cite{snoek2012practical} method to iteratively optimize the proactive model.  In this stage, the proactive model has an acquisition function to guide the selection of the next set of design variable vector. Then the selected data are labeled on the crowdsourcing platform which will be used to optimize the proactive model in the next round.
 
 In the second stage, we build a comparison-tuning model upon the predictive ability of the proactive model. The insight of the comparison-tuning model is that, when users face a single design solution, usually they cannot rate its score confidently, but they can rate which one is better by comparison. In this stage, we aim to build a fine-tuning model to predict the user preference score among the best design solutions.  We first generate a pair of design solutions based on the best design solutions returned by the proactive model, then invite participants to rate which solution is better for each pair. The labeled pairwise comparison data is used to train the comparison-tuning model. 

 FEELER requires several rounds of data labeling on a Baidu's crowdsourcing platform\footnote{https://zhongbao.baidu.com/}. There are 500 participants for labeling the data of FEELER.  Each case was evaluated by at least 20 participants.

\section{Proactive model learning}\label{sec:activelearing}


The first stage of FEELER is to build a proactive model. This stage is an iterative active learning processing. We first generate a batch of design solutions and then invite participants to label their preference score for each solution via a five-point Likert question (1-not at all, 5-Extremely). Then we use the labeled solutions to update the model which is used to guide the generation of design solutions for the next round of data labeling and model learning.  
We can summarize the construction of the proactive model in three steps, which are:
\begin{enumerate}
\item Generating design solutions according to a batch of design variable vectors;
\item Collecting collectively labeled data of all design solutions;
\item Updating the proactive model $f(\cdot)$ and its acquisition function, then generating 
a new batch of design variable vectors, and then go to Step (1).
\end{enumerate}

\begin{figure}[htb]
    \begin{minipage}{0.4\textwidth}
    \centering
    \begin{tabular}{cc}
      \hspace{0mm}\includegraphics[width=0.3\textwidth]{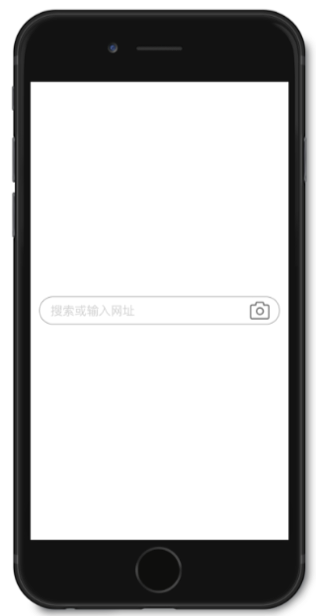} &
      \hspace{-2mm}\includegraphics[width=0.3\textwidth]{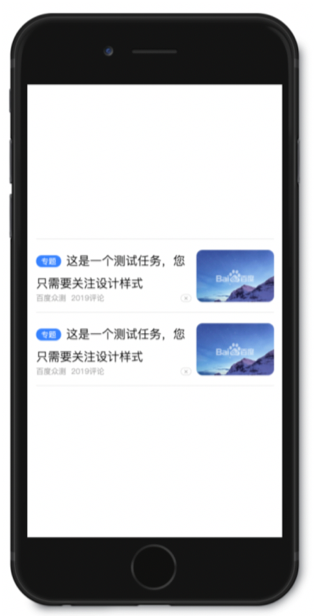} \\
      \hspace{-2mm} (a) Search box &   (b) News feed
   
    \end{tabular}
    \vspace{-3mm}
    \caption{Examples of the testing design solutions.}
    \label{fig:uim_example}
    \end{minipage}
\end{figure}

In the first step, we need to generate a batch of design solutions according to a set of design variable vectors $\mathbf{X}=\{\vec{x}_1, \vec{x}_2, ..., \vec{x}_n\}, \vec{x}_i \in \mathcal{X}$. 
In the initialize step, we generate the design variable vectors by random sampling from the domain $\mathcal{X}$ of the module. In the optimization iteration, the design variable vectors are generated according to an acquisition function in the domain $\mathcal{X}$. We postpone the discussion about the acquisition function to Section \ref{sec:acq}. 
Examples of design solutions of Search Box and News Feed are shown in Figure \ref{fig:uim_example}.
To avoid the influence of other confounding design elements, each interface only contained one design solution which was placed in the center of the screen. 


\subsection{Collective labelling}
In this step, we sent the design solutions to participants to label their preference score on the Baidu crowdsourcing platform.  Participants were required to rate their degree of preference via a five-point Likert question (1-not at all, 5-Extremely). To avoid the bias of a single participant, we send the same design solution $\vec{x}_i$ to $20$ participants,  then we average rating scores of all the participants as the user preference score $y_i$ of the design solution. After the labeling process, we can get a labeled dataset $D=\{(x_i,y_i)\}$. 
Since we adopt an iterative active learning method to label the data, there are multiple rounds to label the solutions. We note the $l$-th round of the labeled data set as $D^l=\{(x^l_i,y^l_i)\}$, $0\leq l\leq L$.

\subsection{Updating the model and acquisition function}\label{sec:acq}
The updating of the proactive model can be explained from the Bayesian optimization perspective. At first we incorporate a prior belief about model $f(\cdot)$. 
After optimizing the $f(\cdot)$ with a labeled data set $D^l$, we can generate another labeled dataset $D^{l+1}$ to optimize $f(\cdot)$ with all the previous labeled data.

Given the labeled data $D^l$, for a new design variable vector $\vec{x}_{t}$, the posterior probability distribution of $f(\vec{x}_t)$ is $P((\vec{x}_t|D^0,..., D^l)$. If we use GPs as the proactive model, then we have $P((\vec{x}_t|D^0,..., D^l)=\mathcal{N}(u(\vec{x}_{t}), \delta^2(\vec{x}_{t}))$ where $u(\cdot)$ and $\delta^2(\cdot)$ are mean and variance matrix defined by Eqn. (\ref{eqn:gp_mean}) and Eqn. (\ref{eqn:gp_var}) respectively. 


After updating the model, we use an acquisition function to guide the selection of candidate design solutions where an improvement over the current best design solution is likely. In other words, we need to identify a set of new design variable vectors that can maximize the acquisition function over $\mathcal{X}$, where the acquisition function is calculated using the updated posterior model. In FEELER we use the Expected Improvement (EI) \cite{mockus1978application} as the acquisition function which can select the design solutions to, in expectation, improve the user preference value upon $f(\cdot)$ the most. For GPs, the analytical expression for $EI(\cdot)$ is:
\begin{align*}
    EI(\vec{x})=\left\{
        \begin{aligned}
       &(u(\vec{x})-f(\vec{x}^*))\Phi(\eta) + \sigma(\vec{x})\phi(\eta)  & if \sigma(\vec{x}) >0 \\
       &max(0, u(\vec{x})-f(\vec{x}^*)) &  if \sigma(\vec{x}) =0
        \end{aligned}
        \right. \\
\end{align*}
where $\phi(\cdot)$ and $\Phi(\cdot)$ denote the probability density function (PDF) and cumulative distribution function (CDF) of the standard normal distribution function, $f(\vec{x}^*)$ is the current best design variable vector, and ${\eta} = \frac{u(\vec{x})-f(\vec{x}^*)}{\sigma(\vec{x})}$.  The advantage of EI is that it can automatically trade off the exploitation versus exploration.
Exploitation means sampling the design variable vector where the $f(\cdot)$ predicts a high value and exploration means sampling design variable vector where the prediction uncertainty (i.e. variance) of $f(\cdot)$ is high.

We use a random sampling method to generate the design variable vectors for the next round of evaluation. We first generate a set of random vectors in the domain of a user interface module, that $H_i=\{\vec{h}_{i,1}, \vec{h}_{i,2}, ..., \vec{h}_{i, a} \}$ and $\vec{h}_{i,j} \in \mathcal{X}$. Then we input the random vectors $H_i$ into the acquisition function $EI(\cdot)$ to select the vector $\vec{h}_{i,*}$ that maximizes $EI(\cdot)$, i.e. $\vec{h}_{i,*} = argmax_{\vec{h}_{i,j}\in H_i}EI(\vec{h}_{i,j})$. Then we take $\vec{h}_{i,*}$ as a candidate design variable vector $\vec{x}_i$. The random sampling process is repeated $b$ times to form a new set of design variable  vectors $X^l=\{\vec{x}^l_1, \vec{x}^l_2, ..., \vec{x}^l_b\}$, where $l$ denotes the order of iteration round. 

\subsection{Lessons and remarks}

There are several issues deserving attention in the proactive model of FEELER. 
The first finding is about the design of the multiple-choice question for crowdsourcing. There are two ways to design the choice question. One is the Yes/No question letting participants indicate whether she/he likes the design solution. The other one is a five-point Likert question (1-not at all, 5-Extremely)
letting participants indicate different levels of his/her preference. We find that Yes/No question is not suitable since most of the participants tend to give a ``No'' answer. One possible reason is that users can always find an unsatisfied point of the user interface module. Thus, we adopt the five-point Likert question. 

Second, some participants may not answer the questions seriously and randomly select a choice. Such behavior will affect the quality of the labeled ground truth. To avoid such a problem, we randomly present duplicate questions to the same participants at different times. If the answers for the same question is quite different (the score difference is larger than 2), we think this participant is unqualified, and remove all her/his answers. If a user always gives extreme choices like 1 or 5, we also remove all her/his answers. The participants did not know such filter rules. 

Third, there is a trade-off to balance the number of labeled solutions and the cost since the larger dataset requires higher cost. In this stage, we design a simple formula to determine the number of labeled instances which is $3\cdot2^{d}$ where $d$ is the dimension number of design variable vector. The intuition of the formula is that we hope there are at least two instances for each dimension, and then we multiply it by $3$ to increase the coverage of the sampled vector over the space.  Therefore, in each round, we generated 1500 ($\approx 3\cdot2^9=1536$) design solutions of Search Box, and 800 ($\approx 3\cdot2^8=768$) News Feed design solutions to be labeled.

\section{Optimizing comparison-tuning model}\label{sec:pl}
\begin{figure}[htb]
    \begin{minipage}{0.5\textwidth}
    \centering
    \begin{tabular}{cc}
      \hspace{0mm}\includegraphics[width=0.45\textwidth]{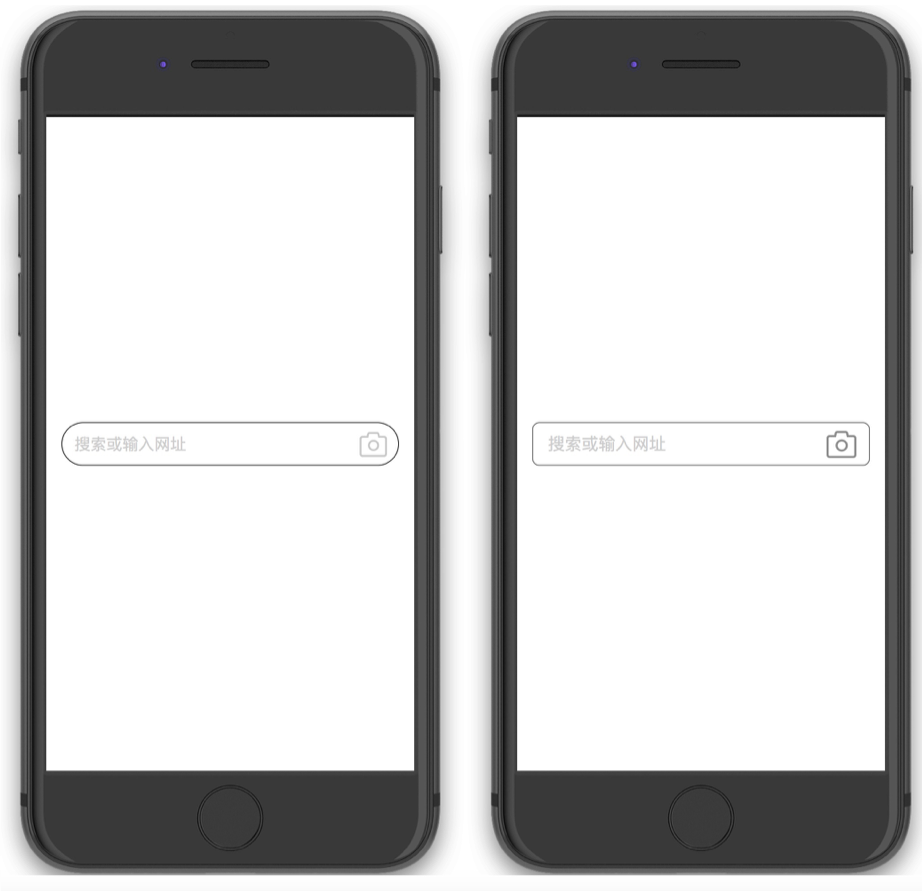} &
      \hspace{-2mm}\includegraphics[width=0.46\textwidth]{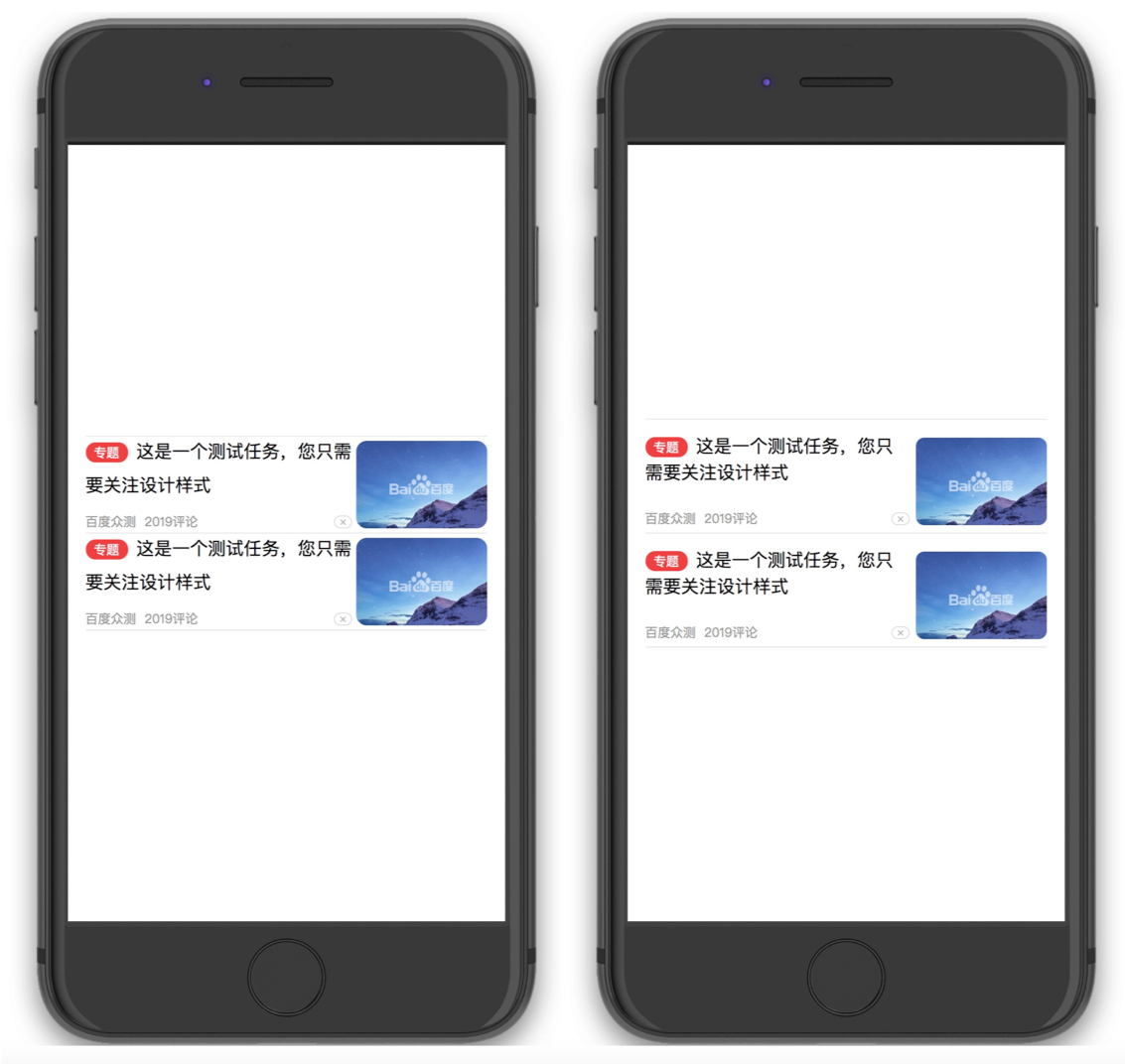} \\
      \hspace{-2mm} (a) Search box &   (b) News feed
    \end{tabular}
     \vspace{-3mm}
    \caption{Examples for solution comparison. Given a pair of design solution, participants rate which one is better.}
    \label{fig:pair_example}
    \end{minipage}
\end{figure}

In the second stage of FEELER,  we build a comparison-tuning model based on the comparison among the best design solutions generated by the proactive model. 
However, the predicted score of the proactive model is based on the five-point Likert question which only reflects the vague subjective judgment for each design solution. In this step, we refine the model by capturing superiority among the best solutions. The main idea is that given the best solutions generated by the previous stage, we randomly select a set of pairs of design solutions, and then invite participants to rate which solution is better. Examples about pairs of design solutions of Search Box and Feed News are illustrated in Figure \ref{fig:pair_example}.

\subsection{Generating candidate solution pairs}
The generation of design solution pairs is based on the proactive model built on the first stage of FEELER. In this step, we first randomly generate a large amount of design solution and then select a set of the best design solutions based on the proactive optimized in Section \ref{sec:activelearing}. Then we randomly construct a set of design solution pairs from the best design solutions set. These solutions pairs are sent to the crowdsourcing platform for preference rating. In our crowdsourcing platform, we use 20 participants to rate each solution pair and determine the preference order by majority voting.  

This process can be formally described as follows. We random generate a large set of design solutions $X'=\{\vec{x}'_1,...,\vec{x}'_N\}$ that $N\to \infty$ (we set $N=30,000$). Given the proactive model $f(\cdot)$, we select a small subset of design solutions $X=\{\vec{x}_1,...,\vec{x}_n\}$($n\ll N$ and $X\subset X'$) that $f(\vec{x}_r)\geq f(\vec{x}'_s)$ if $\vec{x}_r \in X$ and $\vec{x}'_s\in X' \bigcap \vec{x}'_s\notin X$. Then after the collective labeling by the participants on the platform, we can obtain a set of observed comparison labels on the design solution pairs, which can be denoted as $R=\{\vec{x}^m_i\vartriangleright \vec{x}^m_j, m=1,..., M\}$ where $\vec{x}^m_i\in X$ and $\vec{x}^m_j \in X$, and  $\vec{x}^m_i\vartriangleright \vec{x}^m_j$ means design solution $\vec{x}^m_i$ is rated better than solution $\vec{x}^m_j$ voting by participants.



\subsection{Optimization}
Our next objective is to optimize a new comparison-tuning model $g(\cdot)$ which can return preference score for a design solution with the preference relation observed in the data $R=\{\vec{x}^m_i\vartriangleright \vec{x}^m_j, 1\leq m\leq M\}$. Hereafter we refer $\vec{x}_i$ and $\vec{x}_j$ in $R$ with omitting $m$ for simplifying the notation. In this stage, we also assume the comparison-tuning model $g(\cdot)$ as Gaussian Process, and adopt a preference learning method based on GP \cite{chu2005preference,wang2010gaussian,houlsby2012collaborative,wang2014active}.

In order to take into account a measure of the variability in user judgement, we introduce a noise tolerance $\epsilon$ to the comparison-tuning model $g(\cdot)$ which is similar with the 
TrueSkill ability ranking model \cite{herbrich2007trueskill,barber2012bayesian}. The actual score for a solution is $g(\vec{x}_i) + \epsilon_i$ where $\epsilon_i$ is Gaussian noise of zero mean and unknown variance $\delta^2$, i.e. $\epsilon_i \sim \mathcal{N}(\epsilon_i|0, \delta^2)$. The variance $\delta^2$ is fixed across all design solutions and thus takes into account intrinsic variability in the user judgement. Then the likelihood function to capture the preference relation of data $R$ is:
\begin{align}\label{eqn:Px}
  \mathcal{P}(\vec{x}_i\vartriangleright \vec{x}_j|g, \epsilon_i, \epsilon_j)=\left\{
  \begin{aligned}
 &1 & if g(\vec{x}_i)+\epsilon_i > g(\vec{x}_j) + \epsilon_j \\
 &0 & otherwise
  \end{aligned}
  \right.
\end{align}
The marginal likelihood of $\mathcal{P}(\vec{x}_i\vartriangleright \vec{x}_j|g)$ over the Gaussian noise $\mathcal{N}(0,\delta^2)$ is:
\begin{align}
  \mathcal{P}(\vec{x}_i\vartriangleright \vec{x}_j|g) &= \int\int \mathcal{P}(\vec{x}_i\vartriangleright \vec{x}_j|g, \epsilon_i, \epsilon_j)\mathcal{N}(\epsilon_i|0,\delta^2)\mathcal{N}(\epsilon_j|0,\delta^2)\mathrm{d}\epsilon_i \mathrm{d}\epsilon_j.
\end{align}
According to Eqn. \ref{eqn:Px}, we have $\mathcal{P}(\vec{x}_i\vartriangleright \vec{x}_j|g)=\Phi(\frac{g(\vec{x}_i) - g(\vec{x}_j)}{\sqrt{2}\delta})$ where $\Phi(\cdot)$ is the cumulative normal distribution function.

The posterior probability of the comparison-tuning model $g(\cdot)$ is \cite{williams2006gaussian}: 
\begin{align}
\mathcal{P}(g|R) = \frac{\mathcal{P}(g)\mathcal{P}(R|g)}{A},
\end{align}
where $g=[g(\vec{x}_1), g(\vec{x}_2), ..., g(\vec{x}_n)]^T$ and the normalization denominator $A=\mathcal{P}(R)=\int \mathcal{P}(g)\mathcal{P}(R|g)\mathrm{d}g$ is called the marginal likelihood.

We assume that prior probability $\mathcal{P}(g)$ is a zero-mean Gaussian process:
\begin{align}\label{eqn:pg}
  \mathcal{P}(g)=\mathcal{N}(g|{0}, \mathbf{K}),
\end{align}
where the Gram matrix $\mathbf{K}$ (refer to Eqn. \ref{eqn:gp_mean}) is computed over all design solution vector appearing in solution pair data $R$.

The likelihood $\mathcal{P}(R|g)$ is the joint probability of  the observed solution pairs given the model $g(\cdot)$ which is a product of Eqn. \ref{eqn:Px}:
\begin{align}\label{eqn:prg}
\mathcal{P}(R|g)=\prod_{\vec{x}_i\vartriangleright \vec{x}_j \in 
R} \mathcal{P}(\vec{x}_i\vartriangleright \vec{x}_j|g).
\end{align}

The hyper-parameters in the Bayesian framework are the noise variance $\delta^2$ and the kernel width $\Delta$ of RBF kernel. Learning of the hyper-parameters can be formulated as searching optimal values of the hyper-parameters that maximize the marginal likelihood $\mathcal{P}(R)=\int \mathcal{P}(g)\mathcal{P}(R|g)\mathrm{d}g$ which is also called the evidence for the hyper-parameters. However, $\mathcal{P}(R)$ is analytically intractable. There are two categories of methods to solve the marginal likelihood which are 1) approximation method like Laplace approximation \cite{mackay1996bayesian} 
and expectation propagation \cite{minka2001family}; 
and 2) stochastic simulation method like Monte Carlo (MC) Simulation \cite{ferrenberg1989optimized} or Markov Chain Monte Carlo (MCMC) Simulation \cite{carlin1995bayesian}. In this paper, we adopt Laplace approximation in our framework mainly following the method in \cite{chu2005preference}.

The inference learning of the hyper-parameters can be briefly explained as follows, and the detailed explanation can be found in \cite{chu2005preference}.  The posterior probability of $\mathcal{P}(g|R)$ is $\mathcal{P}(g|R) \propto \mathcal{P}(R|g) \mathcal{P}(g)$. Therefore, the maximum a posteriori (MAP) estimation of $g$ (i.e. $g^*=\arg{max}_g\mathcal{P}(g|R)$) appears in the mode  of the following function:
\begin{align}\label{eqn:map}
z(g) = \log(\mathcal{P}(R|g) \mathcal{P}(g))= \sum_{\vec{x}_i\vartriangleright \vec{x}_j \in 
R}\log \Phi(\frac{g(\vec{x}_i) - g(\vec{x}_j)}{\sqrt{2}\delta})-\frac{1}{2}g^T\mathbf{K}^{-1}g.
\end{align} 
Since we have $\frac{\partial z(g)}{\partial g}|_{g=g^*}=0$, Newton method can be used to find the MAP point of Eqn. (\ref{eqn:map}). 

The Laplace approximation of $\mathcal{P}(g|R)$ refers to carrying out the Taylor expansion at the MAP point $g^*$ up to the second order for $z(g)$:

\begin{align}
  z(g) \backsimeq z(g^*) - \frac{1}{2} (g-g^*)^T \Lambda^* (g-g^*),
\end{align}
where $\Lambda^*$ is the Hessian matrix of $-z(g)$ at MAP point $g^*$. $\Lambda^*$  can be re-written as $\Lambda^*=\Omega^* + \mathbf{K}^{-1}$ and $\Omega^*$ is an square matrix whose elements are $\frac{-\partial^2\sum\log \Phi(\frac{g(\vec{x}_i) - g(\vec{x}_j)}{\sqrt{2}\delta})}{\partial g(\vec{x}_i)\partial g(\vec{x}_j)}|_{g=g^*}$. 
Then we have:
\begin{align}\label{eqn:pgr_approx}
  \mathcal{P}(g|R) = exp\{z(g)\} \propto exp\{-\frac{1}{2} (g-g^*)^T \Lambda^* (g-g^*)\},
\end{align}
which means we can approximate the posterior distribution $ \mathcal{P}(g|R)$ as a Gaussian distribution 
with mean as $g^*$ and covariance matrix as $(\Omega^* + \mathbf{K}^{-1})^{-1}$.

By basic marginal likelihood identity (BMI) \cite{chib2001marginal}, we can get the marginal likelihood $\mathcal{P}(R)$ (or evidence) as:
\begin{align}\label{eqn:pr_map}
  \mathcal{P}(R)= \frac{\mathcal{P}(R|g^*)\mathcal{P}(g^*)}{\mathcal{P}(g^*|R)}.
\end{align}
Eqn. (\ref{eqn:pr_map}) can be explicitly computed by combining Eqn. (\ref{eqn:pg}), Eqn. (\ref{eqn:prg}) and Eqn. (\ref{eqn:pgr_approx}). Then we can adopt a gradient descent method to learn the optimal values for the hyper-parameters. 

\subsection{Prediction}

Now given a test design solution $\vec{x}_t$, we would like to obtain its posterior predictive distribution which is:
\begin{align}\label{eqn:pos_pred}
p(g(\vec{x}_t)|R) = \int \mathcal{P}(g(\vec{x}_t)|g)\mathcal{P}(g|R)\mathrm{d}g.
\end{align}

As we have expressed in Eqn. (\ref{eqn:pg}), we assume that $\mathcal{P}(g)$ follows a zero-mean Gaussian Process, then according to Eqn. (\ref{eqn:gp_mean})  and Eqn. (\ref{eqn:gp_var}) we have:
\begin{align}
  \mathcal{P}(g(\vec{x}_t)|g) = \mathcal{N}( g(\vec{x}_t)|\mathbf{k}^T\mathbf{K}^{-1}g, \kappa(\vec{x}_{t},\vec{x}_{t})-\mathbf{k}^{T}\mathbf{K}^{-1}\mathbf{k}).
\end{align}
According to Eqn. (\ref{eqn:pgr_approx}), we can approximate the distribution $\mathcal{P}(g|R)$ as a Gaussian distribution with $g|R \sim \mathcal{N}(g^*, (\Omega^* + \mathbf{K}^{-1})^{-1})$. Thus, the posterior predictive distribution $p(g(\vec{x}_t)|R)$ defined in Eqn. (\ref{eqn:pos_pred}) can be explicitly expressed as a Gaussian $\mathcal{N}(g(\vec{x}_t)|u_t, \delta_t^2)$ with: 
\begin{align}
u_t &= \mathbf{k}^T\mathbf{K}^{-1}g^* \\
\delta_t^2 &= \kappa(\vec{x}_{t},\vec{x}_{t}) - \mathbf{k}^T(\mathbf{K}+\Omega^{*-1})\mathbf{k}
\end{align}
Note that variance  $\delta_t^2$ is simplified as: $(\mathbf{k}^T\mathbf{K}^{-1})^T(\mathbf{K}^{-1}+\Omega^*)(\mathbf{k}^T\mathbf{K}^{-1})+\kappa(\vec{x}_t,\vec{x}_t)-\mathbf{k}^T\mathbf{K}^{-1}\mathbf{k}=
\kappa(\vec{x}_r,\vec{x}_r) -
\mathbf{k}^T(\mathbf{K}^{-1})^T(\mathbf{K}^{-1}+\Omega^*)(\mathbf{K}^{-1})\mathbf{k}=\kappa(\vec{x}_{r},\vec{x}_{r}) - \mathbf{k}^T(K+\Omega^{*-1})\mathbf{k}$.
Thus, given any design solution $\vec{x}_t$ we can compute the posterior predictive distribution of this solution. Usually, we can use the mean $u_t$ as predicted score, and use $\delta_t$ to form confidence intervals.
%

\subsection{Remarks}
Here we discuss several practical issues about the comparison-tuning model. First of all, instead of using the comparison-tuning model directly, we use a two-stage method to learn the oracle model. It is possible to construct the comparison-tuning model without the first stage of FEELER. However, in that case, we will build a model to rank the design solutions in the whole domain. Since there is an almost infinite number of design solutions for each user interface model, building such a model requires a very high labor cost to label the data. It is not practical to obtain a reasonable model in this manner. 

Second, the comparison learning method can achieve better performance than the purely active learning method. Participants can only give a vague judgment about the design solution, whereas they can better capture the slight difference when they compare them in pairs. Our experiments also demonstrate our claim. 

Third, in order to finish the comparison task, all participants were required to conduct this task using mobile phone simulator on PCs. We do try our best to simulate the experience to use the smartphone.

Fourth, in this stage, suppose we select top $n$ best design solutions, we random sample $2\cdot n$ pairs from the best design solutions, i.e. $M=2\cdot n$. In our experiment, we select the best 500 solutions based on the model in the first stage and then generate 1000 pairs to train the model.





\section{Experiments}
We first present the settings as well as experiment evaluations on our method. We also present an in-depth discussion on how to utilize FEELER to quantitatively analyze the design variables.

\subsection{Settings}

\subsubsection{Competitors}
Actually, there is no direct competitor for the exploration of a user interface module. Though some machine learning algorithms can be trained on the dataset generated by FEELER (in two stages),  simply using these competitors cannot solve the user interface module exploration problem. The experiments in this section just verify the predictive capability of FEELER.


Here we use three groups of competitors to evaluate the performance, which includes regression, classification, and learning-to-rank. The first group contains regression models which directly predict the preference score for each design solution, including linear regression (\textbf{LR}) which has good interpretability, Support Vector Regressor (\textbf{SVR})\cite{drucker1997support} which performs well in small datasets and Multilayer Perception Regressor (\textbf{MLPR}) which is a deep learning model with high capacity.  The second group is made up of classification models. We process the preference scores into binary labels by considering a design solution good (labeled as 1) if its preference score higher than 2.5, else it as a bad solution (labeled as 0). Then we implement two binary classifiers as the competitors which are Logistic Regression(\textbf{LogiR}) and Multilayer Perceptron Classifier(\textbf{MLPC}). 
The third group is a learning-to-rank model with assuming there is only one group in the whole dataset. We use the {XGBoost}(\textbf{XGB}) \cite{chen2016xgboost} with setting loss as ``rank:pairwise'' to perform pairwise labeled comparison dataset.  We use Proactive-GP to denote the proactive model built by FEELER in stage one.

\subsubsection{Dataset}
We conduct our experimental evaluations on two datasets generating from Search Box and News Feed of the Baidu App.  The labeled data in the proactive model stage is used as ground truth. Note that such labeled data may not be real ground truth, but can relatively reflect the properties of good design solutions. We randomly split the last round of labeled data in the proactive stage into 80\%, 10\%, and 10\% data as train, validation and test dataset. Then we add the labeled data of all previous rounds into the training data. There are two rounds of data labeling in stage one in our experiment. Search Box has 1500 while News Feed has 800 labeled instances in each round. For the comparison-tuning model (of FEELER) and XGB (of the learning-to-rank model), we use the same 1000 solution pairs to train the model while the testing set is the same with other baselines.

\subsubsection{Metrics}\label{pra:Metrics}
Here we adopt the Average Precision (AP) and Normalized Discounted Cumulative Gain (NDCG) as the metrics \cite{li2011short}. To calculate the metrics, we rank all design solutions by their preference score labeled by participants, and then we sort all the predicted results of models above to obtain predicted rankings. Please refer to Appendix \ref{apx:metrics} about the description of the metrics. By default, we set the default threshold $\rho$ of AP as 0.1 and the default fold number $n$ in NDCG as 15.


\subsection{Performance evaluation}

\begin{table}[tbh]
\vspace{-4mm}
\caption{Performance comparison on AP and NDCG.} \centering
\vspace{-4mm}
\begin{tabular}{c|c|c|c|c}
  \hline
  Dataset &  \multicolumn{2}{|c|}{Search Box} &\multicolumn{2}{|c}{News Feed}\\\hline
  Model & AP & NDCG & AP & NDCG\\\hline \hline
  \textbf{FEELER} & \textbf{0.226} & \textbf{0.668} & \textbf{0.275} & \textbf{0.673}\\\hline
  Proactive-GP & 0.167  & 0.648 & 0.129 & 0.544\\\hline
  LR & 0.185  & 0.604 & 0.156 & 0.578\\\hline
  SVR & 0.159  & 0.570 & 0.117 & 0.474\\\hline
  MLP-R &  0.182  & 0.590 & 0.134 & 0.526\\\hline
  LogiR & 0.164  & 0.588 & 0.156 & 0.556\\\hline
  MLP-C & 0.149  & 0.576 & 0.154 & 0.525\\\hline
  XGB & 0.181  & 0.599 & 0.130 & 0.515\\\hline
\end{tabular}
\label{tab:AP_NDCG}
\vspace{-3mm}
\end{table}


Table \ref{tab:AP_NDCG} shows the prediction performances of FEELER and its competitors on AP and NDCG metrics. 
As we can see, FEELER achieved higher AP and NDCG than other models on both the Search Box dataset and the News Feed dataset. 
Moreover, FEELER could do a better job than Proactive-GP, which demonstrates the effectiveness of our second stage to build a comparison-tuning model. We also evaluate the performance of proactive-GP by comparing with other regression models under Mean Absolute Error in Appendix \ref{apx:proactive}.
Note that FEELER can not only predict the preference score for each design solution but also conduct variable analysis which is discussed in Section \ref{sec:analysis}. 


\begin{figure}[htb]
    \begin{minipage}{0.5\textwidth}
    \centering
    \begin{tabular}{cc}
      \hspace{0mm}\includegraphics[width=0.45\textwidth]{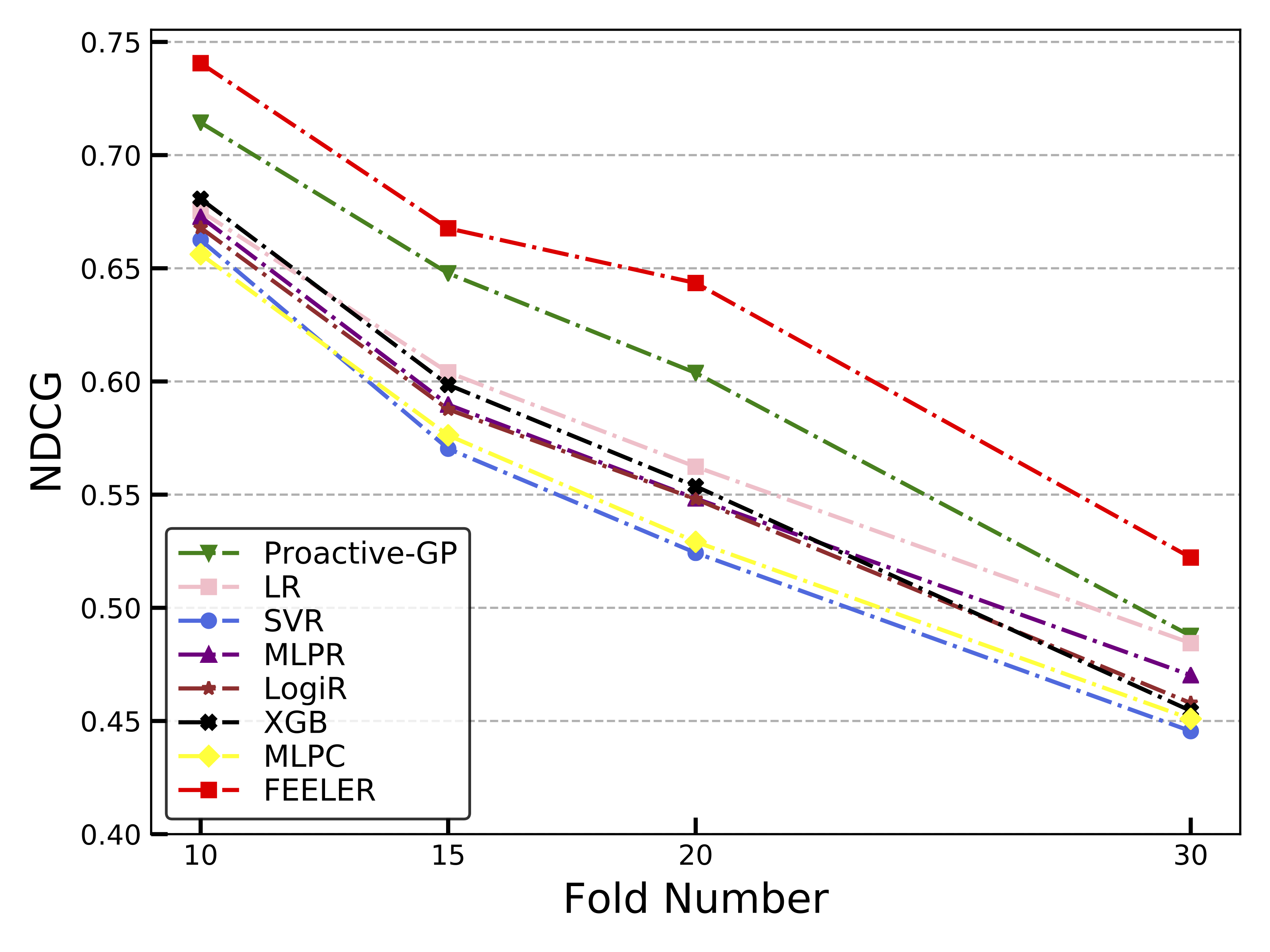} &
      \hspace{-2mm}\includegraphics[width=0.46\textwidth]{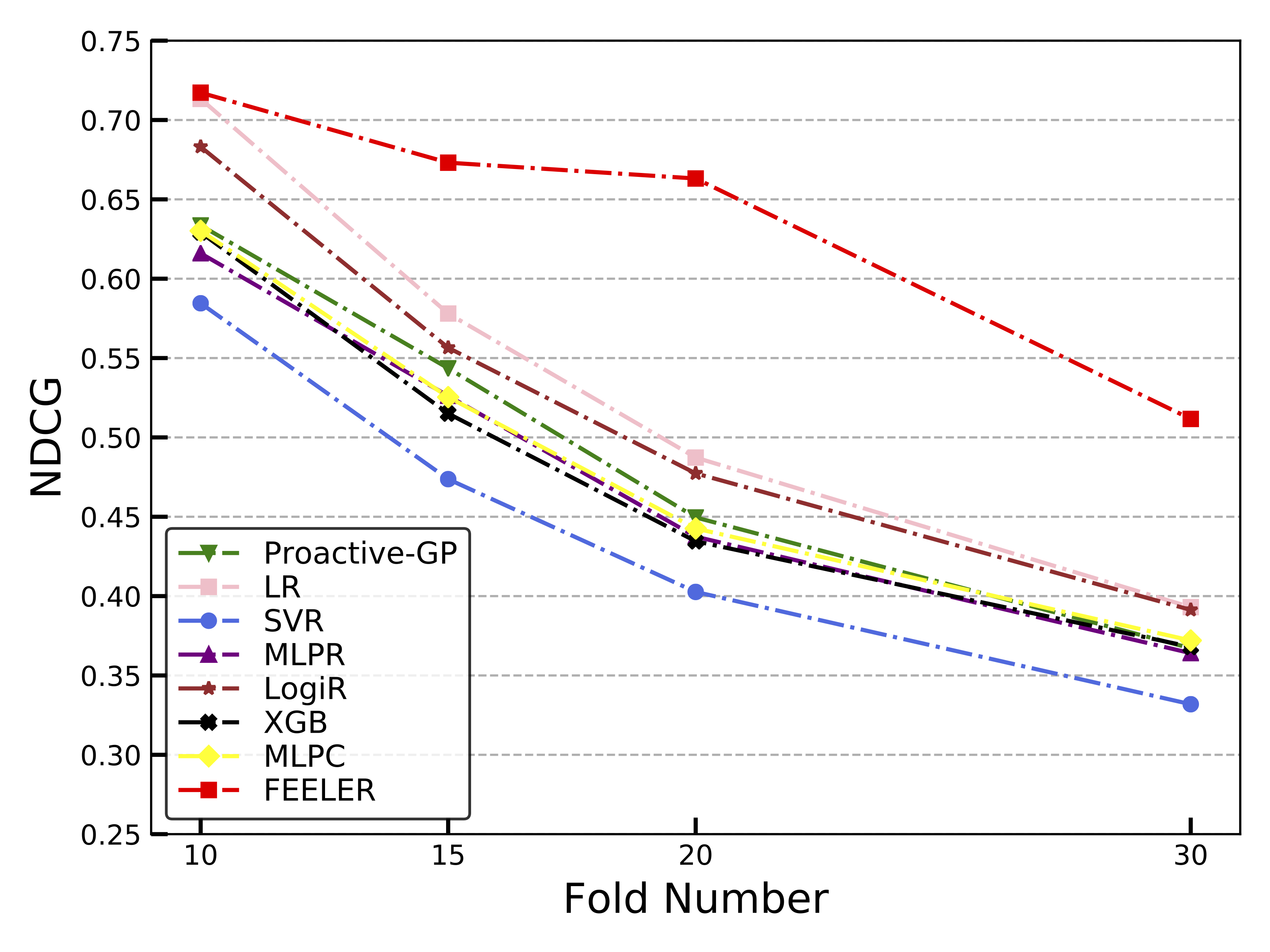} \\
      \hspace{-2mm} (a) Search box &   (b) News feed

    \end{tabular}
    \vspace{-3mm}
    \caption{NDCG with varying fold number $n$.}
    \label{fig:ndcg}
    \end{minipage}
\end{figure}






Figure \ref{fig:ndcg} shows the NDCG  with different fold number $n$. 
As we can see from Figure \ref{fig:ndcg}, all the competitors declined drastically with the increasing of fold number $n$ since they could not rank the solutions properly, because larger fold number means a more strict condition for correct ranking. Meanwhile, the NDCG of FEELER is always larger than all competitors, meaning FEELER can make a better prediction with fine-grained ranking. This is especially useful for user interface design since we care more about how to find the best design solutions from good solutions. 



\subsection{Utilization of FEELER for variable analysis}\label{sec:analysis}
The most important application of FEELER is to predict the preference score given a design solution. Using our developed tool, the designers of the Baidu App can adjust different design variables to see the trend of preference score.
Moreover, FEELER also provides a mechanism to quantitatively analyze the design variables. We discuss this in this section. 


\subsubsection{Distribution of top design solutions}



\begin{figure}[htb]
    \begin{minipage}{0.5\textwidth}
    \centering
    \begin{tabular}{cc}
      \hspace{-3mm}\includegraphics[width=0.45\textwidth]{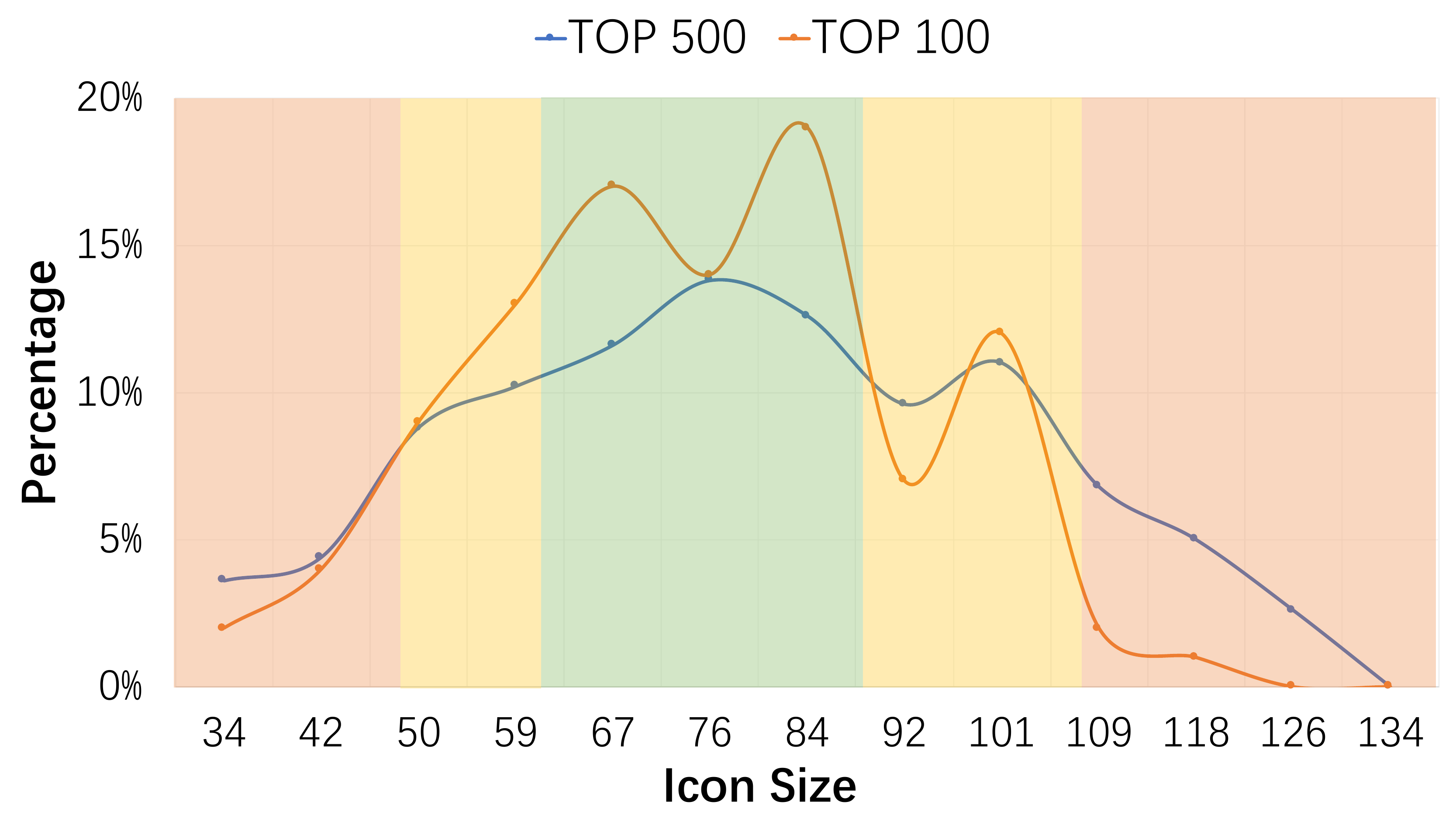} &
      \hspace{-2mm}\includegraphics[width=0.45\textwidth]{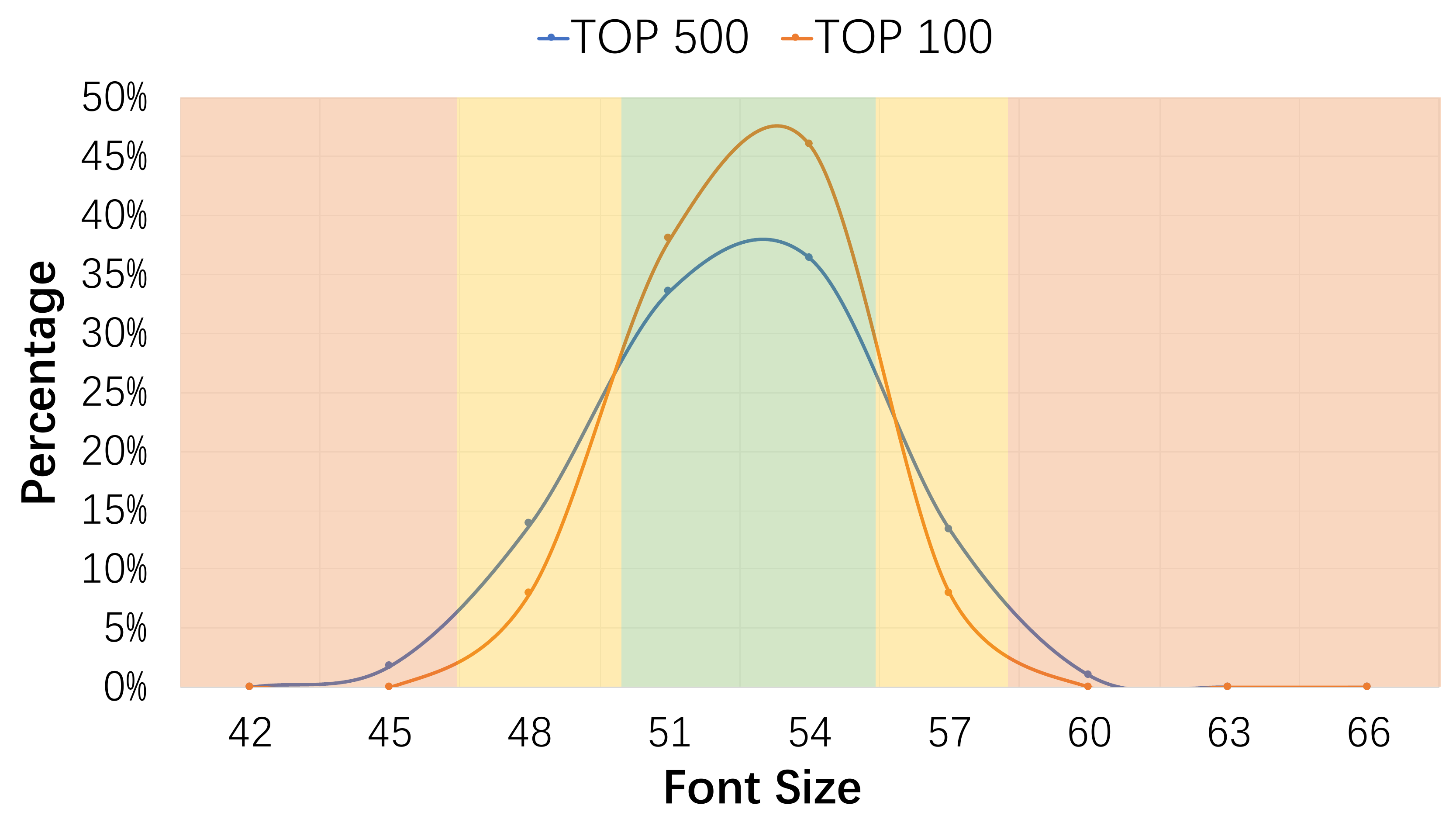} \\
      \hspace{-2mm} (a) $Icon$ $Size$ of Search Box &   (b) $Font$ $Size$ of  News Feed
    \end{tabular}
    \vspace{-3mm}
    \caption{Distribution of  top design solutions.}
    \label{fig:var_dis}
    \end{minipage}
\end{figure}

We first showcase the relationship between the preference score and the design variables by calculating the distribution of top design solutions. To conduct such analysis, we randomly generate 30,000 design solutions for Search Box and News Feed respectively and then use FEELER to predict their score. 
Then we select the top 500 and top 100 design solutions  with the highest score. Figure \ref{fig:var_dis}(a) shows the distribution under the design variable $icon$ $size$ for Search Box; and Figure \ref{fig:var_dis}(b) shows the distribution under the design variable $font$ $size$.
From both figures, we can find that the distribution of Top500 solutions and Top100 solutions are almost consistent with similar peak values. (The distribution of Top100 are more concentrated.) 
Figure \ref{fig:var_dis} can also help us determine the best values for each design variables. The green area in Figure \ref{fig:var_dis} is the proper range of the design variables given by designers. We can see that most of the good design solutions are within such given intervals. Moreover, we can also find the best value (i.e. peak value in Figure \ref{fig:var_dis}) of design variables that has the largest chance to get the highest preference score. Such peek values for these design variables are unknown by the designer.



\subsubsection{Multivariate density distribution of design variables}

\begin{figure}[htb]
    \begin{minipage}{0.5\textwidth}
    \centering
    \begin{tabular}{cc}
      \hspace{0mm}\includegraphics[width=0.45\textwidth]{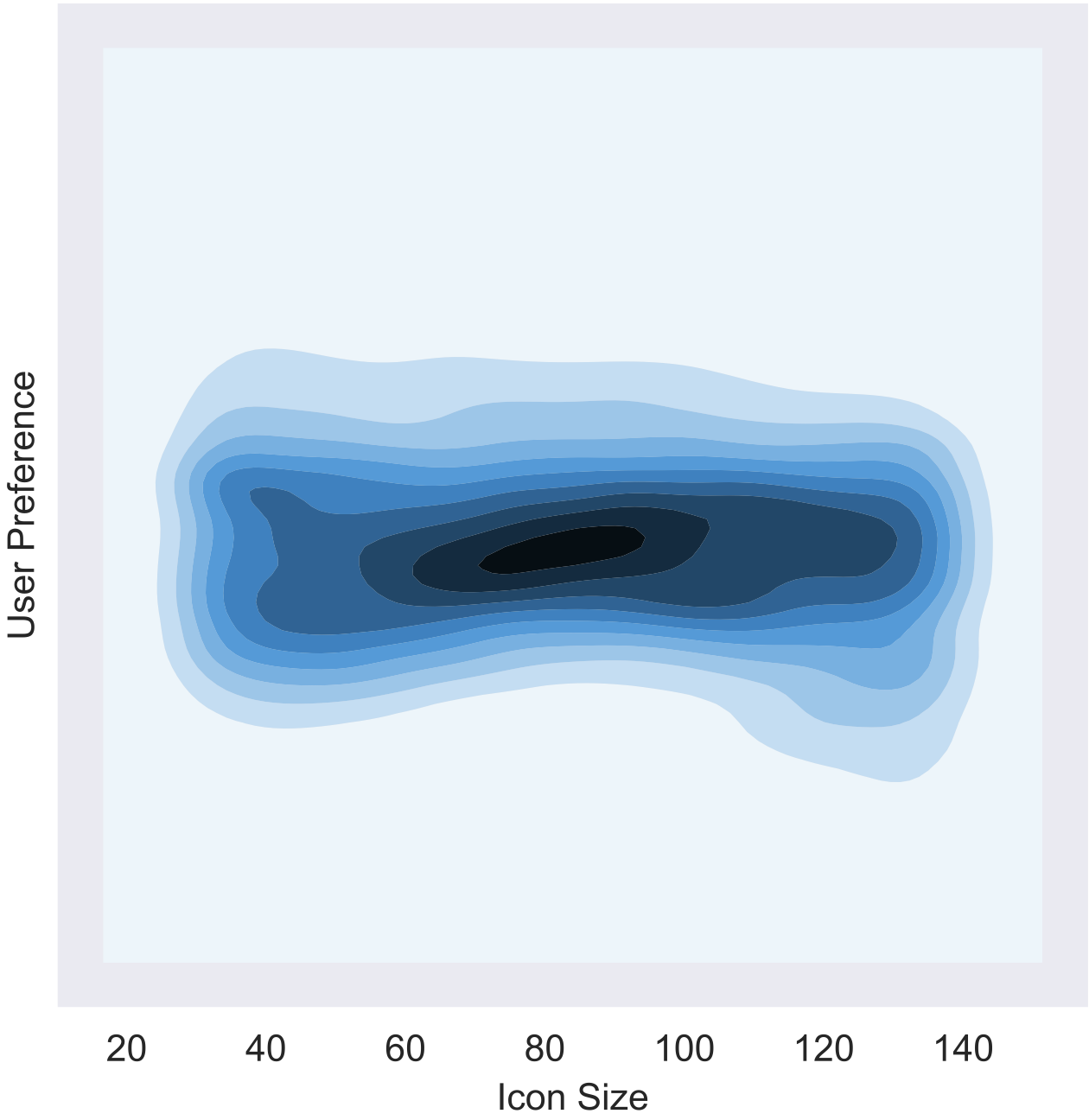} &
      \hspace{-2mm}\includegraphics[width=0.45\textwidth]{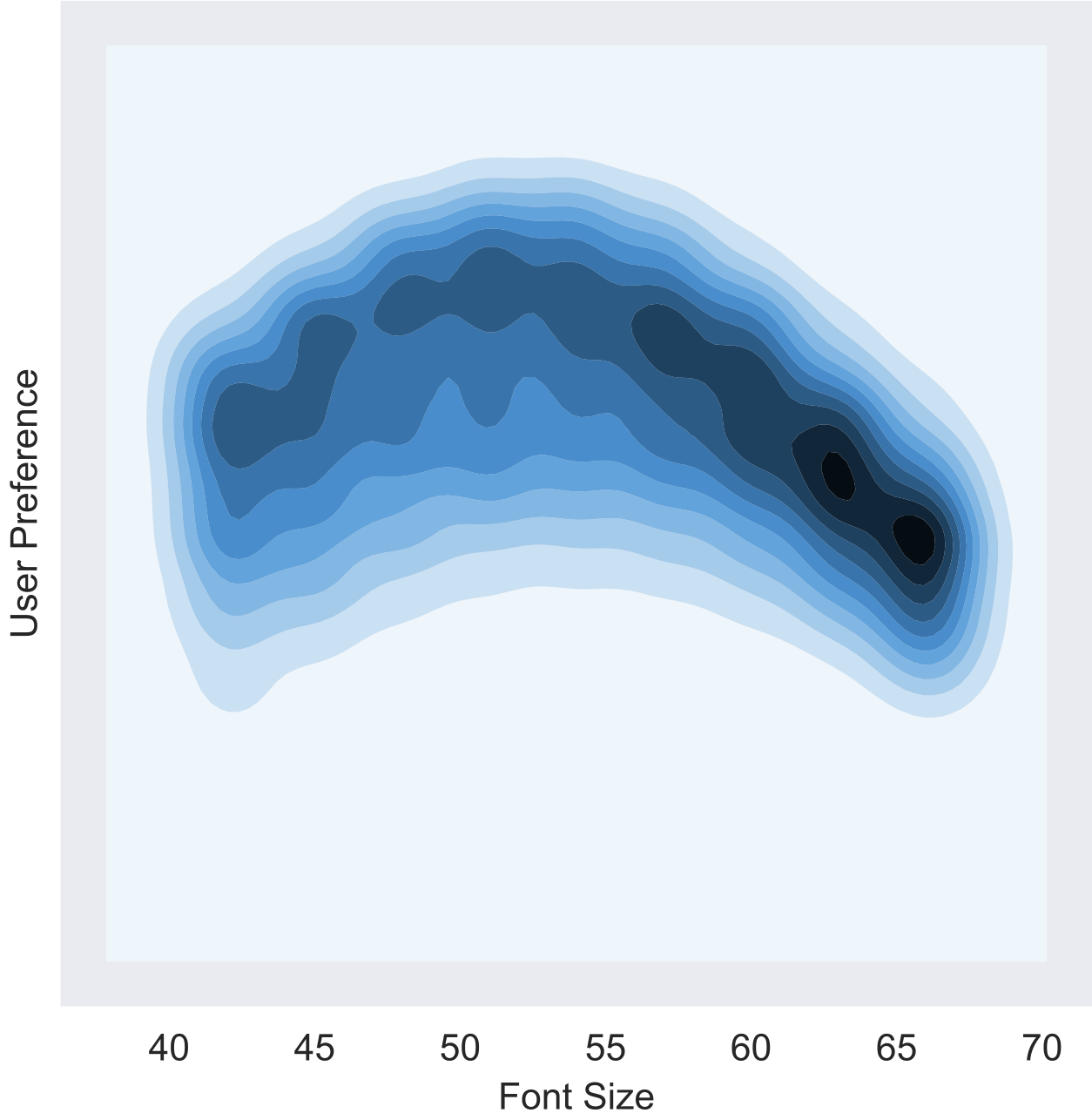} \\
      \hspace{-2mm} (a)  &   (b) 
    \end{tabular}
    \vspace{-3mm}
    \caption{Multivariate density distribution with varying design variables. (a): Search box with varying $Icon$ $Size$; (b): News Feed with varying $Font$ $Size$.}
    \label{fig:kde}
    \end{minipage}
\end{figure}

Since FEELER is a statistical model, we can build the multivariate density distribution of design variables to show the correlation distribution between preference score and design variables. Figure \ref{fig:kde} shows such distribution on Search Box($icon$ $size$ vs preference score) and News Feed ($font$ $size$ vs preference score). By the multivariate density distribution, we can analyze the effect of a single variable on the preference score. For example, as shown in Figure \ref{fig:kde}, with varying the design variables ($icon$ $size$ and $font$ $size$), the probability density distribution of preference score is changed. We can also see that $font$ $size$ of News Feed has a larger impact on the probability density distribution than the one of $icon$ $size$ of Search Box. 

\subsubsection{Variable correlation analysis}
\begin{figure}[htb]
    \begin{minipage}{0.5\textwidth}
    \centering
    \begin{tabular}{cc}
      \hspace{0mm}\includegraphics[width=0.48\textwidth]{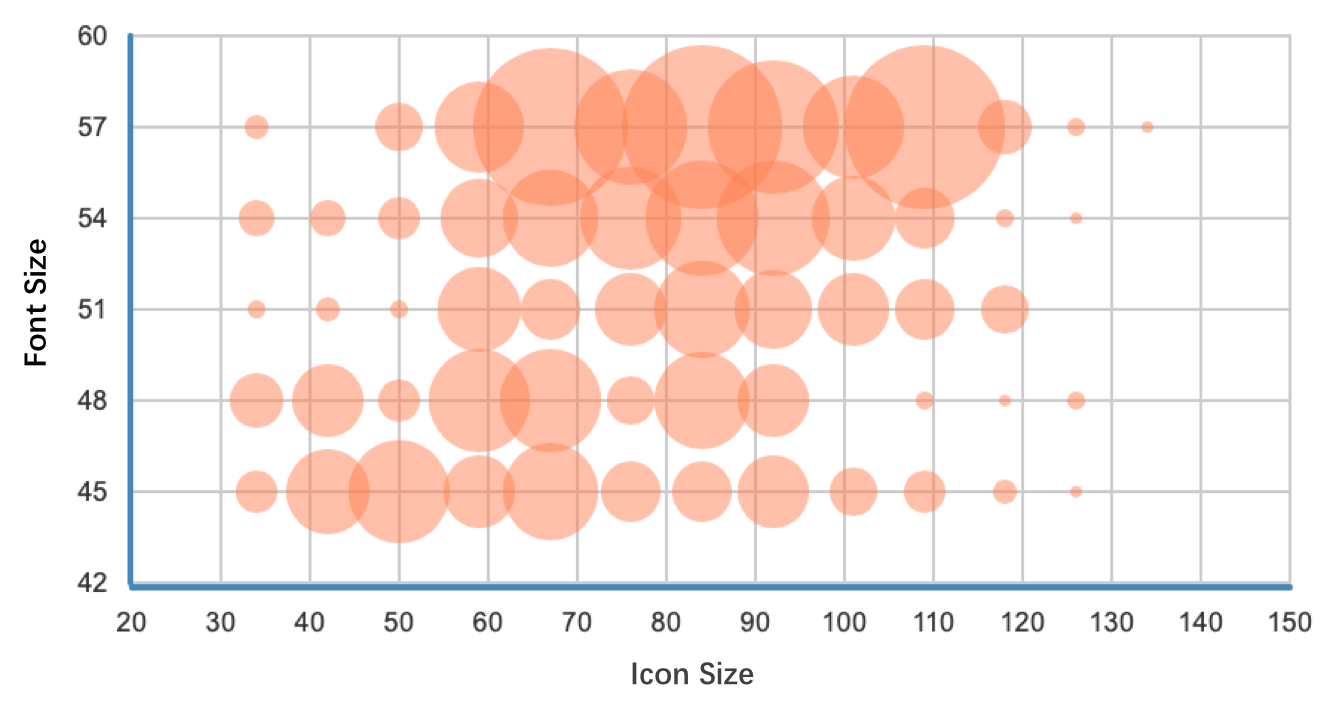} &
      \hspace{-2mm}\includegraphics[width=0.48\textwidth]{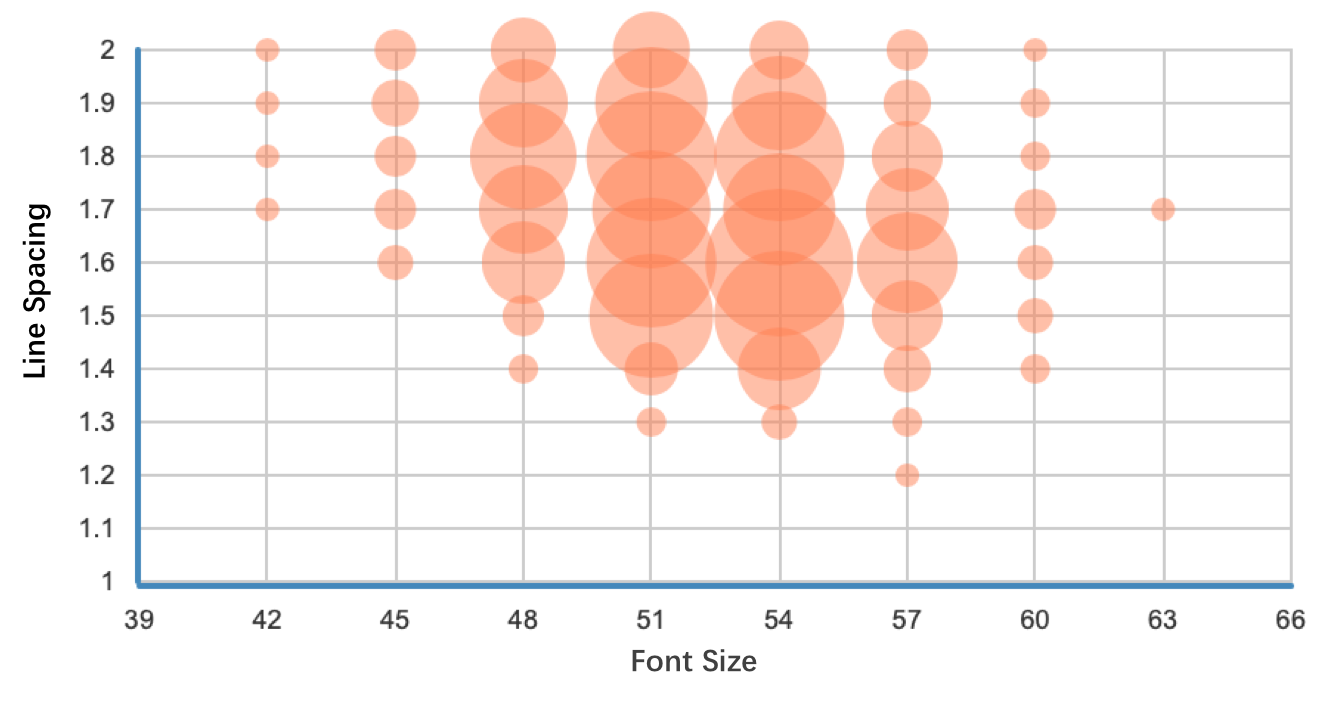} \\
      \hspace{-2mm} (a) Search Box  &   (b) News Feed

    \end{tabular}
    \vspace{-3mm}
    \caption{Variable correlation on Top500 design solutions. (a)$Icon$ $Size$ vs $Font$ $Size$ of Search Box; (b) $Font$ $Size$ vs $Line$ $Spacing$ of News Feed.}
    \label{fig:2var}
    \end{minipage}
\end{figure}

FEELER can also help us to observe the interaction relations between design variables.
Figure \ref{fig:2var}  shows the joint distribution of two design variables of Top500 design solutions of Search Box and News Feed via bubble diagram. In both figures, the larger the bubble in the figure, there are more design solutions with the design variables being indicated by the bubble. Therefore, the bubble diagram figures can help us observe the correlation among design variables, which can help designers make decisions. For example, assume the designer has fixed $icon$ $size$ as $85px$ for Search Box, the best range of $font$ $size$ for Search Box should be about $55px$ to $58px$. Using FEELER, designers could easily choose proper values for design variables.

\section{Related work}

There are only a few existing works related to our paper.
In \cite{seguin2019triptech}, the authors propose a solution to evaluate the design concept of a product which is quite different from a user interface. However, the proposed method in \cite{seguin2019triptech} is solely based on survey data and no machine learning methods are discussed. 
Authors in \cite{kayacik2019identifying} also discuss how to design a user-friendly machine learning interface with the user experience and research scientist collaboration. It is also a survey-based method without utilizing any machine learning technology. A mobile interface tappability prediction method had been investigated recently \cite{swearngin2019modeling}, but this method does not touch the user interface module design problem.
There are also some recent studies to predict touchscreens tappability \cite{swearngin2019modeling} and accessibility \cite{guo2019statelens}, but these methods usually make predictions based on existing screens, without investigating how to help designers optimize and generate design solutions. In recent years, there are also some works to utilize the machine learning and deep learning to model and predict human performance in performing a sequence of user interface tasks such as menu item selection \cite{li2018predicting}, game engagement \cite{khajah2016designing,lomas2016interface} and  task completion time \cite{duan2020optimizing}. To the best of our knowledge, there is no existing work to utilize machine learning to assist the user interface design 
of mobile App with collective learning.

\section{Conclusion, lessons and What’s next}

We investigated to explore the best design solution for a user interface module of a mobile app with collective learning. 
FEELER collects user feedback about the module design solution in a process of multiple rounds, where a proactive model is built with active learning based on Bayesian optimization, and then a comparison-tuning model is optimized based on the pairwise comparison data. Thus, FEELER provides an intelligent way to help designers explore the best design solution for a user interface module according to the user preference.  FEELER is a statistical model with Gaussian Processes that can not only evaluate design solutions and identify the best design solutions, but also can help us find the best range of design variables and variable correlations of a user interface module.  FEELER has already been used to help the designers of Baidu to improve the user satisfaction of the Baidu App, which is one of the most popular mobile apps in China.

There are several lessons learned from our method. First of all,  machine learning methods can help us to identify the best design solution for a user interface module, which shed some light on a new machine learning-based user interface design paradigm. Second, FEELER can also help designers to understand the hidden rules for good design solutions of a user interface module. We can use FEELER to identify the impact of a single factor and reasonable range of design variables without having to exhaustively manually evaluate all the design solutions.

We will continue to extend FEELER to be a general tool to improve the user satisfaction of other mobile apps. 
Moreover, it also deserves research attention to investigate how to apply the methodology of FEELER to generate the whole user interface of a mobile App, which is a more challenging problem due to the complexity of the user interface.




\begin{acks}
This research is supported in part by grants from the National Natural Science Foundation of China (Grant No.71531001,61972233,U1836206).
\end{acks}

\bibliographystyle{ACM-Reference-Format}
\bibliography{feeler_ref}

\newpage
\appendix

\section{More experiments}
\subsection{Metrics}\label{apx:metrics}
Average Precision (AP) is employed to evaluate the performance of ranking. 
Given an user-labeled ranking and a predicted ranking, we could compute AP by:
\begin{align}\label{eqn:AP}
AP = \frac{1}{K}{\sum_{i=1}^{N}}(\frac{s_{i}}{i}(\sum_{j=1}^{i}s_{i})),
\end{align} 
where $N$ is the number of design solutions in user-labeled ranking as well as predicted ranking. We set the score value $s$ of top $K=\rho N (0<\rho<1)$($\rho$ is called threshold) design solutions in user-labeled ranking data as 1 and otherwise 0. In the predicted ranking, for the $i$-th solution we obtain its score value $s_i$ according to its score in the original rank in user-labeled ranking data.


Normalized Discounted Cumulative Gain (NDCG) is another metric. To compute NDCG, we cut the user-labeled ranking data into $n$ folds, each fold contains $m$ design solutions. We mark the scores of $m$ design solutions in $j$-th fold as the same, which is $S=n-j+1$. 
For each design solution in the predicted ranking data, we obtain its score value $s_i$ according to its score in the original rank in user-labeled ranking data.
In this way, we can compute NDCG by:

\vspace{-2mm}
\begin{align}\label{eqn:DCG}
DCG = \sum_{i=0}^{n\times m}\frac{2^{s_i}-1}{log_2(i+1)}
\end{align} 

\vspace{-2mm}
\begin{align}\label{eqn:IdealDCG}
IdealDCG = \sum_{i=0}^{n\times m}\frac{2^{S_i}-1}{log_2(i+1)}
\end{align} 

\vspace{-2mm}
\begin{align}\label{eqn:NDCG}
NDCG = \frac{DCG}{IdealDCG}
\end{align} 

By default, we set the $\rho$ of AP as 0.1 and the $n$ of NDCG as 15, which are selective enough without losing too much variety. 

We also use Mean Absolute Error (MAE) to evaluate the performance of score predicting by regression model:
\vspace{-1mm}
\begin{align}\label{eqn:MAE}
MAE = \frac{1}{N}{\sum_{i=1}^{N}}|label_{i}-pred_{i}|,
\end{align} 
where $N$ is the number of design solutions in testing set while $label_i$ and $pred_i$ are the actual score and prediction score of $i$-th design solution respectively.

\subsection{Evaluation of Proactive-GP}\label{apx:proactive}

\begin{table}[tbh]
\vspace{-4mm}
\caption{Performance comparison on MAE.} \centering
\vspace{-4mm}
\begin{tabular}{c|c|c}
  \hline
  Model/Datasets &  \multicolumn{1}{|c|}{Search Box} &\multicolumn{1}{|c}{News Feed}\\\hline \hline
  {Proactive-GP} & {0.476} & {0.208}\\\hline
  LR & 0.424  & 0.233\\\hline
  SVR & 0.437  & 0.244\\\hline
  MLPR &  0.427  & 0.222\\\hline
\end{tabular}
\label{tab:MAE}
\vspace{-4mm}
\end{table}
We evaluate the performance of Proactive-GP by comparing it with other regression models under MAE. As shown in Table.\ref{tab:MAE}, Proactive-GP achieved 0.476 and 0.206 Mean Average Error(MAE) on Search Box dataset and News Feed dataset respectively.  Proactive-GP does not always have the smallest MAE on all datasets compared with baselines. It is because the main objective of Proactive-GP is to balance the exploitation versus exploration trade-off, whereas accurate prediction is not its main optimization task.






\end{document}